\documentclass[pdflatex,sn-mathphys]{sn-jnl}

\usepackage{tabularx}

\jyear{2021}%

\theoremstyle{thmstyleone}%
\theoremstyle{thmstyletwo}%

\theoremstyle{thmstylethree}%

\begin{document}

\title[Minecraft Elasticity]{Elasticity Solver in Minecraft for Learning Mechanics of Materials by Gaming}


\author[1]{\fnm{Zachariah} \sur{Beck}}\email{becks@purdue.edu}
\equalcont{These authors contributed equally to this work.}

\author[1]{\fnm{Brandon} \sur{Alpert}}\email{balpert@purdue.edu}
\equalcont{These authors contributed equally to this work.}

\author[3]{\fnm{Alexander} \sur{Bowman}}

\author[3]{\fnm{William R.} \sur{Watson}}

\author*[1,2]{\fnm{Adrian} \sur{Buganza Tepole}}\email{abuganza@purdue.edu}

\affil[1]{\orgdiv{Mechanical Engineering}, \orgname{Purdue University}, \orgaddress{ \city{West Lafayette},  \state{Indiana}, \country{USA}}}

\affil[2]{\orgdiv{Weldon School of Biomedical Engineering}, \orgname{Purdue University}, \orgaddress{ \city{West Lafayette},  \state{Indiana}, \country{USA}}}

\affil[3]{\orgdiv{Department of Education}, \orgname{Purdue University}, \orgaddress{ \city{West Lafayette},  \state{Indiana}, \country{USA}}}


\abstract{Video games have emerged as a medium for learning by creating engaging environments, encouraging creative and deep thinking, and exposing learners to complex problems. Unfortunately, even though there are increasing examples of video games for many basic science and engineering concepts, similar efforts for higher level engineering concepts such as mechanics of materials are still lacking. Here we present a mesh-free elasticity solver implementation in the popular video game Minecraft, a sandbox game where players can build any structure they can imagine. Modifications to the game, called \textit{mods} in the Minecraft community, are a common feature of this platform. Our elasticity \textit{mod} computes the stress and deformation of arbitrary structures and colors the blocks with a heat-map to visualize the result of the analysis. We used this \textit{mod} in the Honors section of two courses taught at Purdue University:  Basic Mechanics I Statics, Mechanics of Materials. This \textit{teaching tip} describes our experience developing and deploying this tool to encourage its use in biomedical engineering classrooms. A future goal is to engage  the broader audience Minecraft players that already interact regularly with \textit{mods}.}




\maketitle

\section{Challenge Statement}\label{sec1}

Video games have become a medium for learning \cite{squire2008video}. Recent studies support that video games used in educational settings have potential for promoting motivation, complex problem solving, and deep comprehension, especially within a problem-based learning (PBL) framework \cite{watson2012pbl}. Increased technical competency has been shown when students are encouraged to do active, peer-to-peer, and self-learning \cite{rhoads2014purdue}, which are characteristic in video game environments \cite{coller2014learning}. Still, there are few video game tools for solid mechanics concepts \cite{coller2007implementing}. Two notable related examples are the \textit{Truss Me} and \textit{Poly bridge} games which are truss solvers in 2D \cite{rimoli_2014,drycactuslimited_2015}. The challenge we tackle is the lack of interactive and engaging learning environments for solid mechanics concepts within a PBL framework. The video gaming space is prime for this challenge because games are already a common interest for the US population: 90\% of the American youth (less than 18 years-old) and 70\% of those aged 18-34 noted that they play video games according to a Pew survey \cite{brown_2020}. Minecraft is a popular sandbox game modeled after the real world, where players can use \textit{blocks} made out of different \textit{materials} to build all kinds of structures\cite{mojang_2022}. Due to its open-ended design which allows players to explore freely and create whatever they can imagine, Minecraft has not only amassed a staggering number of users, but has even led to the stand alone Minecraft Education Edition, a version of the game specifically targeted for teachers to create lessons on basic science and engineering concepts \cite{bar2020crafting}. Unfortunately, many of the efforts to create gaming experiences in Minecraft are limited to basic science concepts \cite{short2012teaching}. On the other hand, one of the unique features of Minecraft is the ability to create modifications to the game, also called \textit{mods}. We take advantage of this flexibility and we show an implementation of a version of smooth particle hydrodynamics (SPH) for solids \cite{bonet2004variational,rausch2017modeling}. We created lessons using this \textit{mod} which were preliminary tested in the Honors section of Basics Mechanics I Statics, and Mechanical of Materials at Purdue University.

\section{Novel Initiative}

\begin{figure}[h!]
\centering
\includegraphics[width=0.9\linewidth]{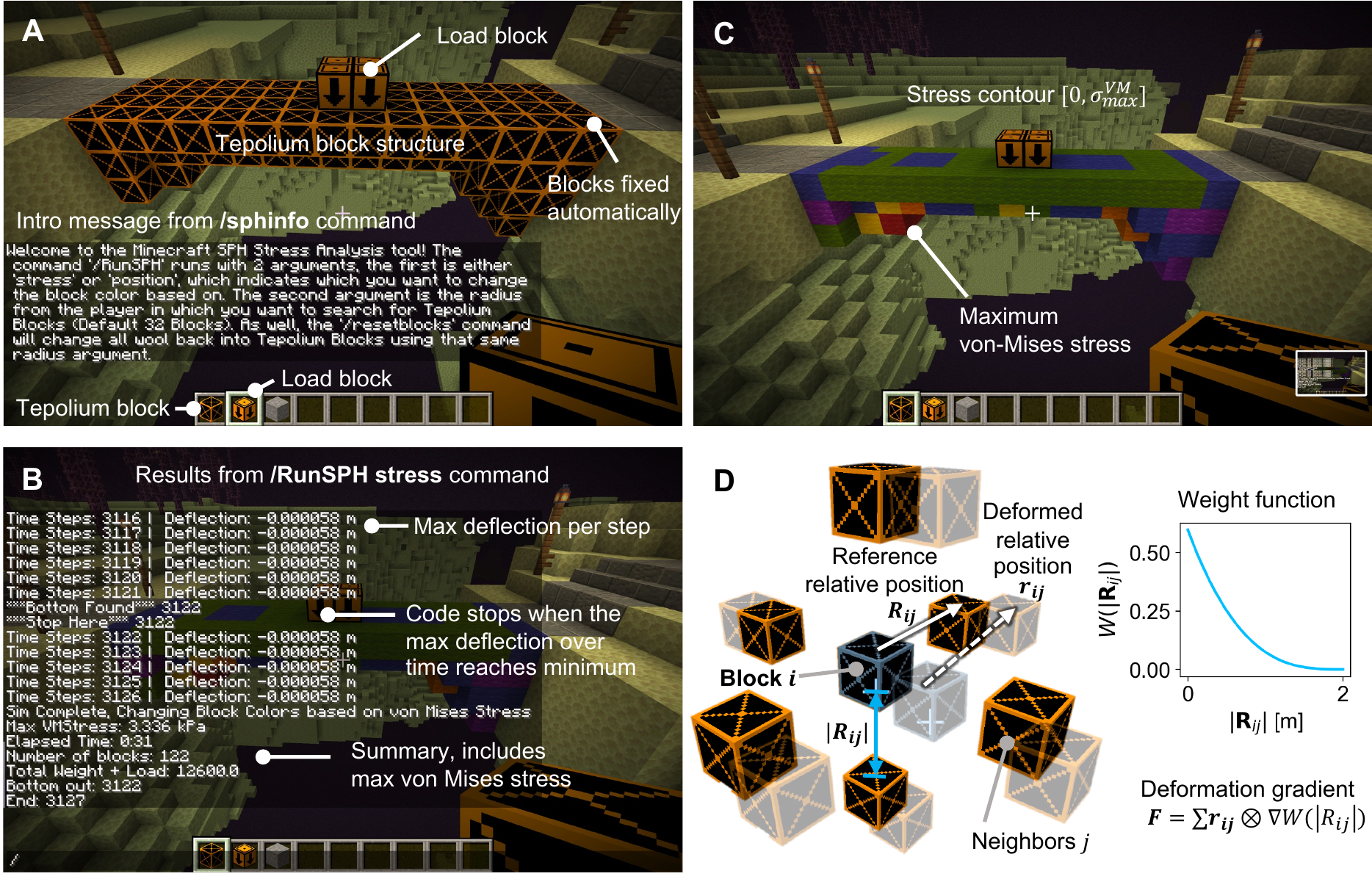}
\caption{ A) Creating structures in Minecraft with the \textit{Tepolium} block. B) Running the \textit{mod} with the \texttt{RunSPH} command identifies all the blocks of the structures and their neighbors, integrates the elastodynamic equations in time, and outputs the deflection of the center of mass to the screen as well as the maximum stress at the end of the simulation. C) Blocks are switched to \textit{wool} of different colors to indicate the amount of stress between 0 and the maximum value of von-Misses stress. D) The core of the algorithm is the interpolation of the deformation gradient based on the initial and final relative positions between particle $i$ and its neighbors $j$ through the weight function $W$. }
\label{Figure:fig1}
\end{figure}

To create engaging learning environments for solid mechanics concepts, we leverage the flexibility of Minecraft \textit{mods} and recent advances in mesh-free methods for linear and nonlinear elasticity. The elasticity solver follows recent particle-based solves, akin to SPH but for solids \cite{bonet2004variational,rausch2017modeling}. This framework is ideal for Minecraft because the world in this video game is already discretized in terms of particles or blocks, placed in a voxelized grid. We created a special material in the game, called \textit{Tepolium}, such that structures can be defined by identifying the \textit{Tepolium} blocks and their neighbors. A sample structure is shown in Fig. \ref{Figure:fig1}A. We deal with boundary conditions and loading in an intuitive way by making blocks in contact with the ground fixed and creating special load blocks, also depicted in Fig. \ref{Figure:fig1}A. Running the \textit{mod} in the game using the \texttt{RunSPH} command computes the displacement and stress distributions from solving the corresponding elastodynamic equations (see Section \ref{sec3}). The output to the command line is depicted in Fig. \ref{Figure:fig1}B, and the heat-map of von-Misses stress in Fig. \ref{Figure:fig1}C. Statics and mechanics of materials classes are core to most biomedical engineering and mechanical engineering programs. We deployed the \textit{mod} in a small class setting to test and refine the problems that we plan to incorporate in future offerings, when we will also thoroughly test the effectiveness of this approach to impact student learning. 

\subsection{Minecraft Elasticity Solver}\label{sec3}

For the reader interested in the implementation, some of the basic details of the solver are covered here, but more extensive discussions on SPH are available in \cite{liu2010smoothed}. Additionally, the Github link with our implementation within the Minecraft Java editor \textit{MCreator} \cite{klemen_matej_2022} is listed at the end of the \textit{Teaching tip}. 

Consider a particle $i$ with $j$ nearest neighbors up to some cutoff $h$. The original positions of these particles are $\mathbf{X}_i$, $\mathbf{X}_j$, and the deformed positions are $\mathbf{x}_i$, $\mathbf{x}_j$. The the deformation gradient is 

\begin{equation}
    \mathbf{F} = \sum \mathbf{r}_{ij} \otimes \tilde{\nabla} W(\|\mathbf{R}\|_{ij}) 
    \label{eq:F}
\end{equation}

where $\mathbf{r}_{ij}$, $\mathbf{R}_{ij}$ are relative position vectors as illustrated in Fig. \ref{Figure:fig1}D, and $W$ is an interpolation function. We remark that because of the Minecraft discretization, all particles have initial volume $V_i=1$ m\textsuperscript{3} and the volume is thus omitted as opposed to, for example \cite{lluch2019breaking}. For the interpolation function we use 

\begin{equation}
    W(\|\mathbf{R}\|_{ij})  = \frac{15}{\pi h^6}(h-\|\mathbf{R}_{ij}\|)^3
    \label{eq:W}
\end{equation}

The gradient in eq. (\ref{eq:F}) does not directly correspond to the gradient of the function in eq. (\ref{eq:W}). Rather, to satisfy that $\mathbf{F}$ is the identity at the initial state, a correction to the gradient is needed. The tensor

\begin{equation}
    \mathbf{A} = \sum \mathbf{R}_{ij} \otimes {\nabla} W(\|\mathbf{R}\|_{ij}) 
    \label{eq:A}
\end{equation}

is computed with the gradient $\nabla W$ from direct partial differentiation of eq. (\ref{eq:W}). Then, $\mathbf{A}$ can be used to get the corrected gradient 

\begin{equation}
    \tilde{\nabla} (\circ) = \nabla(\circ) \mathbf{A}^{-1} \, .
    \label{eq:nabla_tilde}
\end{equation}

Given the deformation gradient for a particle using eq. (\ref{eq:F}), and a choice of constitutive model, the second Piola Kirchhoff stress tensor $\mathbf{S}$ can be calculated. For example, for a St. Venant Kirchhoff solid we have

\begin{equation}
    \mathbf{S} =  \lambda_{VK} \mathrm{tr}(\mathbf{E})\mathbf{I} + 2\mu_{VK}\mathbf{E}\, ,
\end{equation}

with $\mu_{VK}$, $\lambda_{VK}$ material parameters, and $\mathbf{E}=(\mathbf{F}^T\mathbf{F}-\mathbf{I})/2$ the Green-Lagrange strain tensor. The internal forces are actually computed with the first Piola Kirchhoff stress

\begin{equation}
    \mathbf{P} = \mathbf{F}\mathbf{S} + 2J\eta (\bar{\mathbf{d}}\mathbf{F}^{-1})
\end{equation}

where we have additionally considered a viscous damping term depending on the parameter $\eta$ proportional to the isochoric symmetric velocity gradient $\bar{\mathbf{d}}$ for improving convergence to equilibrium and avoid harmonic oscillations. Given the stress for a particle and its neighbors, the net internal force in a particle is 

\begin{equation}
    \mathbf{F}_{\mathrm{int}} = \sum_j (\mathbf{P}_i + \mathbf{P}_j)\tilde{\nabla}W(\|\mathbf{R}_{ij}\|)
\end{equation}

We also consider stabilization terms as suggested in \cite{lluch2019breaking,rausch2017modeling,liu2010smoothed}. 

Finally, given position, velocities and accelerations at time $t$, the explicit time integration scheme for each particle is

\begin{equation}
\begin{aligned}
    \mathbf{V}_{1/2} &=   \mathbf{V}_{t} + \frac{\Delta t}{2}\mathbf{A}\\
    \mathbf{x}_{t+\Delta t} &=  \mathbf{x}_{t} + \Delta t\mathbf{V}_{1/2}\\
    \mathbf{A}_{t+\Delta t} &= \frac{1}{m}(\mathbf{F}_{\mathrm{int}}+\mathbf{F}_{\mathrm{ext}})\\
    \mathbf{V}_{t+\Delta t} &= \mathbf{V}_{1/2} + \frac{\Delta t}{2} \mathbf{A}_{t+\Delta t}
\end{aligned}
\label{eq:integrate_time}
\end{equation}

Where $\mathbf{F}_{\mathrm{ext}}$ are external forces which can be from load blocks or from the weight of \textit{Tepolium} blocks. Boundary conditions are enforced after each time step update from eq. (\ref{eq:integrate_time}).

As mentioned before, the command \texttt{RunSPH} runs the program. All the commands and their description are listed in Table \ref{tab1}.

\begin{table}[h]
\begin{center}
\caption{Commands in Minecraft Elasticity}\label{tab1}%
\begin{tabular}{p{0.18\linewidth} p{0.16\linewidth} p{0.56\linewidth} }
\toprule
Command & Arguments  & Description \\
\midrule
\texttt{/SPHinfo}    & none   & Prints basic information of the mode to screen.  \\
\texttt{/RunSPH}    & \textit{stress}/\textit{position} radius   & Runs the elasticity solver. The first argument has to be either \textit{stress} or \textit{position} and specifies the coloring of the blocks at the end of the simulation. The second argument is an integer. Only particles within a given radius will be checked to initialize the structure.  \\
\texttt{/SPHproperties}    & none/parameter value  & If no argument is passed it prints the list parameters of the simulation and their current values. To modify a parameter, then the first argument is the name of the parameter and the second argument its new value. For example, to change the ultimate stress to 4kPa type: \texttt{SPHproperties ult\_stress 4.0}  \\
\texttt{/resetblocks}&none&Resets the wool blocks back to \textit{Tepolium}.\\
\texttt{/setSpecialBlock}&CoordX CoordY CoordZ&Instead of printing out the deflection of the centroid, which is the default, it prints the deflection of the block at location (CoordX,CoordY,CoordZ) which are integers that must be passed as arguments.\\
\botrule
\end{tabular}
\end{center}
\end{table}

\subsection{Examples}\label{subsec2}

The Minecraft solver was used to create activities for the Honors version of ME270 Basic Mechanics Statics and ME323 Mechanics of Materials at Purdue University in Fall 2021. These courses have a standard curriculum but students have the option to do extra activities to earn the Honor designation for the course. These additional activities usually require a 1 hour weekly workload on top of the standard course. For the students enrolled in the Honors version of these two classes in Fall 2021, the weekly activities consisted of using the Minecraft solver to carry out design challenges and then writing short reports describing how they did the design and to justify with concepts learned in class why their designs worked. We show two examples here. 

\subsubsection{Desert Bridge Challenge}\label{subsubsec2}

\begin{figure}[h!]
\centering
\includegraphics[width=0.9\linewidth]{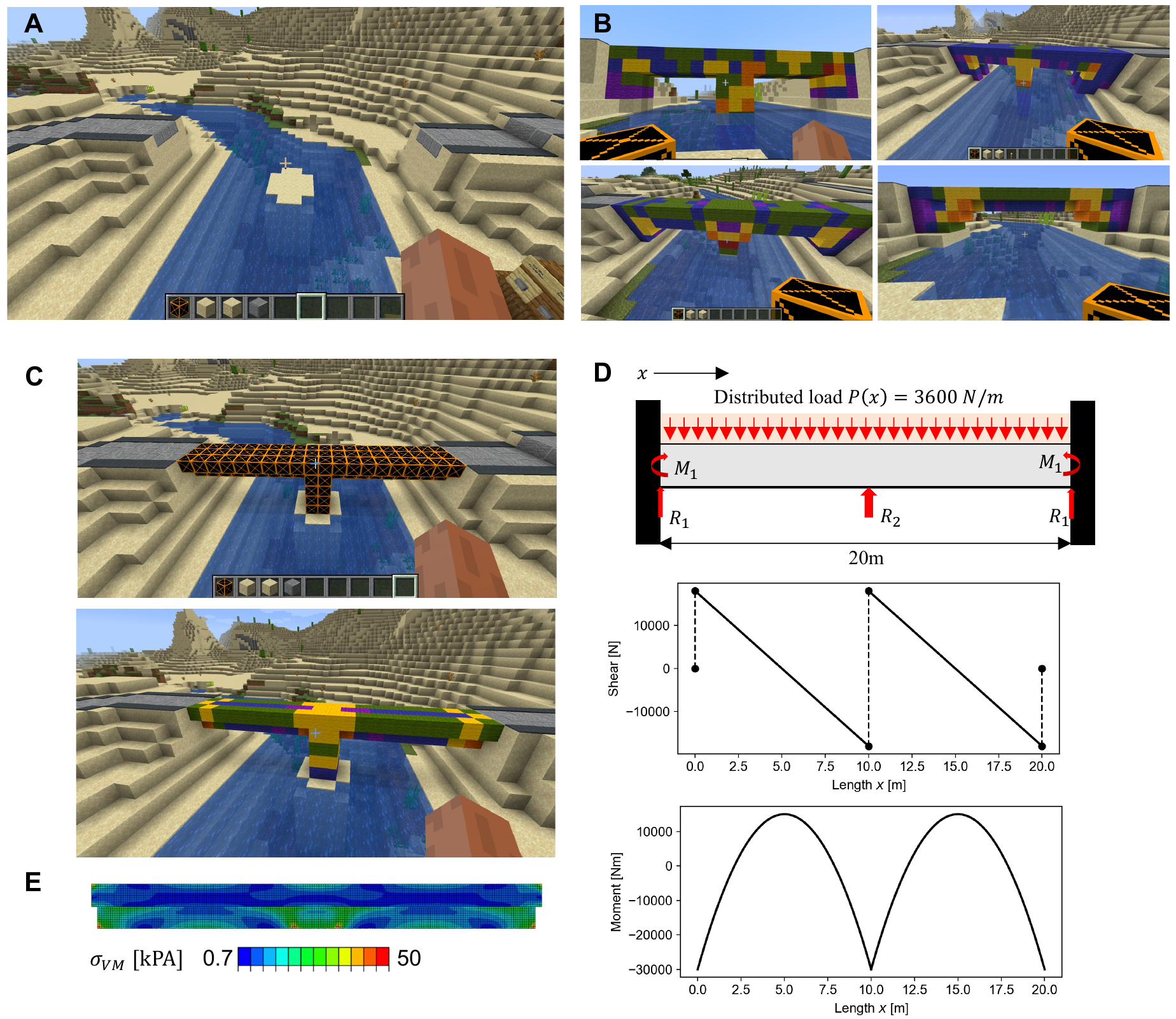}
\caption{ A) Desert bridge challenge setup. B) Example bridge designs created by students that satisfy the design criteria. C) Instructor's solution. D) Analysis of the structure using shear force and bending moment diagrams. E) Finite element solution of the problem using Abaqus. }
\label{Figure:fig2}
\end{figure}

Consider the problem in Fig. \ref{Figure:fig2}A. The prompt to students is to design a bridge across a desert road over a river oasis. The span of the bridge is $20$ m, and the width is $3$ m. The weight of each block is $-900$ N. The constraint is that the stress cannot exceed $15$ kPa as predicted in the game. The Young's modulus of the material is $E=1\times10^9$ Pa, and the Poisson ratio is $\nu =0.4$. Students are encouraged to design a bridge that minimizes the number of blocks to use. 

The problem is designed such that a straight beam connecting the two ends of the road will exceed the $15$ kPa limit. Students were encouraged to try the simplest solution in case it would work. The next, most intuitive, solution, is to add a support at the middle. Fig. \ref{Figure:fig2}B shows feasible designs submitted by the students. At the point of the course at which we did this challenge the students had not encountered shear force and bending moment diagrams. The justification from the students was based on intuition rather than a rigorous analysis. However, we attach a handout with the analysis of our design as a Supplemental material. The details of our solution are illustrated in  Fig \ref{Figure:fig2}C-E. The instructor solution in Minecraft is depicted in Fig. \ref{Figure:fig2}C, the corresponding shear and bending moment analysis is shown in \ref{Figure:fig2}D, and a finite element simulation for the problem is shown in Fig. \ref{Figure:fig2}E. Without a formal understanding of stresses in beams, students intuitively understand that they need to minimize the material in the floating regions compared to the fixed support regions; they also intuitively maximize the bending stiffness by placing blocks preferentially in plane orthogonal to the bridge horizontal plane. It should be noted, as explained in the Supplement, that the elasticity solver in Minecraft is a coarse solution and that finer discretizations are needed to achieve more accurate solutions.   

\subsubsection{Cross-section Challenge}\label{subsubsec3}

\begin{figure}[h!]
\centering
\includegraphics[width=0.99\linewidth]{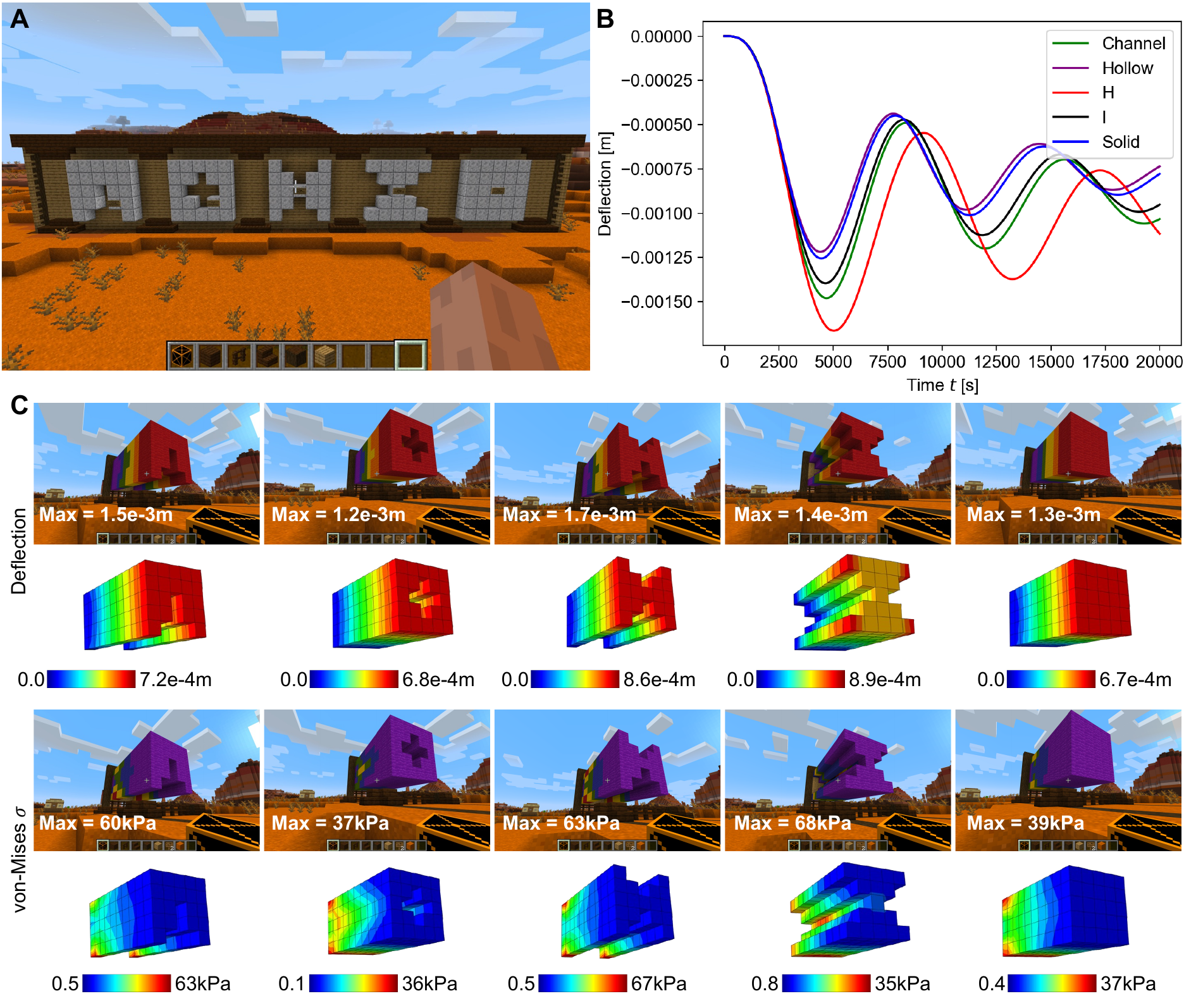}
\caption{ A) Cross section design challenge is to create beams of 10 m with different cross section to understand how moment of inertia affects deflection and stress. b) The deflection over time at the tip of the beam for the different cross sections shows the \textit{H} cross section to be the one with the largest deflection, as expected. C) The contour plots for deflection and stress for each of the beams. The deflection contours are scaled with respect to the maximum of each beam, whereas the stress contours are with respect to the maximum allowed stress of $65$ kPa. }
\label{Figure:fig3}
\end{figure}

A follow up to the study of the desert challenge is the cross section challenge world illustrated in Figure \ref{Figure:fig3}A. The students are given this prompt: We want to create cantilever beams with the cross sections shown, the beams will span 10 m and we are interested in the deflection at the end. Rank them in order of most to least deflection. 

Then the students submit their answers and have to build and test their assumptions. The winner gets a prize at the end of the session (some chocolates in our case). Students are allowed to work in groups to submit their answers and to test out the result. The deflection over time for a block at the tip of the each beam is plotted in Figure \ref{Figure:fig3}B. The mod outputs a \textit{*.csv} file with the displacement and stress of the \textit{SpecialBlock}. The curves in \ref{Figure:fig3}B illustrate that the code is actually integrating dynamic equations and that some oscillations are expected. The results seen in Minecraft are shown in Figure \ref{Figure:fig3}C, which also contrasts the Minecraft solution against a finite simulation with the same level of coarseness in the mesh. The maximum deflection is larger in Minecraft because, as seen in Figure \ref{Figure:fig3}B, the solver overshoots the amount of deformation initially. Nevertheless, the ordering of the beams in terms of deflection can still be achieved with the \textit{mod}. The contours of von Mises stress are consistent between the \textit{mod} and the finite element result with the exception of the \textit{I} beam. The handout solution is given as a Supplement. While there are some innacuracies in the Minecraft solver, students build up the intuition that the \textit{H} beam deflects the most.  

\section{Implementation}\label{sec12}

Watson and Fang \cite{watson2012pbl} developed a framework for structuring a game-based lesson within a PBL framework. This section contains summarized strategies relevant to the five main components identified by the authors: \textit{Problem}, \textit{Activation}, \textit{Exploration}, \textit{Reflection}, \textit{Facilitation}. This serves as a brief tutorial; readers are encouraged to read the full paper for more details. 

Because games are inherently problem based, aligning a lesson with the \textit{Problem} component generally entails checking the compatibility of lesson goals and the game, as the learning should come from both the game and the lesson. For \textit{Activation}, connect gameplay events or outcomes with students’ prior knowledge; when students have intuitive understanding, helping them identify/label the knowledge will help them mentally organize the formal instruction. A successful \textit{Exploration} component should be structured as an iterative trial-and-error process in which students are able to analyze the problem and generate solutions freely, but the learning goals should be clearly delineated. Because many games do not provide opportunities for \textit{Reflection}, the instructor will need to add them in. Ideally, such moments would involve peer collaboration and should be structured around moments in which students discover new data related to the problem, the solution, their investigative processes, or their own knowledge and skills relevant to the learning objective. For \textit{Facilitation} of the lesson, instructors are encouraged to use open-ended questioning strategies that help students articulate their reflections by connecting them to the gameplay experience. Successful \textit{Facilitation} might also include more direct summarization, coaching, or modeling. 

Table \ref{tab2} offers examples of how an instructor could structure their lesson using this particular Minecraft \textit{mod} and the desert bridge challenge.  

\begin{table}[h]
\begin{center}
\caption{Examples of Mod Implementation}\label{tab2}%
\begin{tabular}{p{0.14\linewidth} p{0.76\linewidth} }
\toprule
Component & Example \\
\midrule
Problem & Creating a bridge aligns with lesson goals and with gameplay. The lesson (supports reduce internal shear and moments and hence maximum stress) was reinforced by the gameplay (the bridge would change colors when unsupported blocks exceeded 15 kPa of stress).    \\
Activation    & Students will intuitively support the middle of the bridge. Encourage them to explain their prior knowledge in terms relevant to the game and the lesson. For example, they might restate a simple phrase like “I tried to hold up the middle” to “I placed a vertical support where the stress was the highest”  \\
Exploration & The game structure affords students the freedom to learn through trial and error and provides (mostly) instantaneous feedback on stress and strain. Students were encouraged to use as few blocks as possible, but it was made clear that 15 kPa was a hard limit on stress.   \\
Reflection  & In this implementation, students were allowed to share with each other while playing. Afterwards, a whole-class discussion helped students formalize their insights, and the instructor summarized key takeaways. If teaching virtually, instructors might also employ reflective journaling and screencasts.  \\
Facilitation    & After initial prompts, students freely explored as instructors asked questions like “why did you employ that particular design?” A struggling student might be told “try reinforcing more at the base and see what happens” and a student who solves it quickly might be asked "what can you do to reduce the number of blocks required?” \\
\botrule
\end{tabular}
\end{center}
\end{table}

\section{Reflection}\label{sec13}
This teaching tip shows the use of video games to teach mechanics concepts. Video games are emerging tools for learning \cite{watson2012pbl,watson2016games}, yet there are not many examples of video games that can be leveraged for applying advanced engineering topics at the undergraduate level with a few exceptions \cite{coller2007implementing,coller2009video}. We chose Minecraft in particular because it is already a very popular game and because it allows for a large degree of flexibility through the implementation of modifications or \textit{mods}.  Hence, we reasoned that it would be a good platform to implement advanced concepts while at the same time allow us to reach a wide audience, potentially beyond the classroom environment. For the slightly more formal route of incorporating video games in the classroom within a PBL framework we deployed the elasticity solver as the honors project of ME270: Statics, and ME323: Mechanics of Materials at Purdue University in Fall 2021. The Statics course is handled by the School of Mechanical Engineering, but it serves other departments, such as Biomedical Engineering. We only had 13 students enrolled, plus two undergraduate students helping with the developing of the activities (these two students are co-authors in this paper). The students appreciated the use of Minecraft because they had been familiar with it before, but we also had students who had never played Minecraft before but that were excited about learning while playing this video game. We had 9 men and 4 women in this class, which is a higher ratio than our general ratio of women in the department of Mechanical Engineering (17\% women), and a smaller ratio of women compared to the School of Biomedical Engineering (50\% women). Additionally, the students in the Honors version of a class tend to be already good learners. Thus, using Minecraft in this small group was useful to refine our activities and get initial feedback, but we look forward to a formal evaluation of the use of our elasticity solver to improve learning of mechanics concepts in a future study.  

\backmatter

\bmhead{Supplementary information}

Mod source code: 
Materials available at Tepole lab: \url{https://engineering.purdue.edu/tepolelab/minecraft}

\bmhead{Acknowledgments}

This work was supported by NSF award CMMI 1911346.



\end{document}